# Statistical Kinetics of Phase-Transforming Nanoparticles in LiFePO$_4$ Porous Electrodes

Peng Bai* and Guangyu Tian

*State Key Laboratory of Automotive Safety and Energy, Department of Automotive Engineering,*

*Tsinghua University, Beijing 100084, People's Republic of China*

*Corresponding Author: pengbai@mit.edu

Present Address: Department of Chemical Engineering, Massachusetts Institute of Technology, 77

Massachusetts Avenue, Cambridge, MA 02139, USA

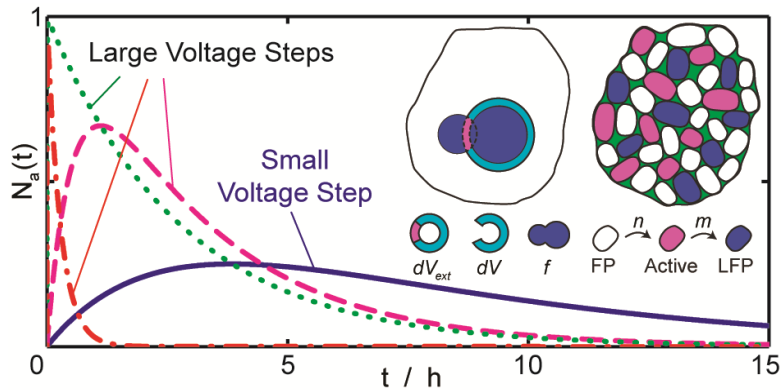

**ABSTRACT**: Using a simple mathematical model, we demonstrate that statistical kinetics of phase-transforming nanoparticles in porous electrodes results in macroscopic non-monotonic transient currents, which could be misinterpreted as the nucleation and growth mechanism by the Kolmogorov-Johnson-Mehl-Avrami (KJMA) theory. Our model decouples the roles of nucleation and surface reaction in the electrochemically driven phase-transformation process by a special activation rate and the mean particle-filling speed of active nanoparticles, which can be extracted from the responses of porous electrodes to identify the dynamics in single composing nanoparticles.

# 1. Introduction

Kinetics of lithium intercalation and phase-transformation is one of the most intriguing puzzles that have attracted intensive investigations since the introduction of LiFePO$_4$ [1]. It is also a critical dynamic factor limiting the high-rate performance of nano-LiFePO$_4$ electrodes. Although the phase boundary propagation in a single particle can be directly observed by the state-of-the-art *in situ* techniques [2], how the phase transformation process is influenced by the statistical effects caused by countless composing particles is yet to be addressed. Neglecting statistical effects, in interpreting the macroscopic performance of porous electrodes as the microscopic behavior of a single particle, or extrapolating battery performance from single particle properties, will result in discrepancies of the basic physics in both scales [3, 4].

Given the technical challenges in conducting single-particle experiments, it is much easier to apply voltage steps to a porous electrode and infer the phase transformation pathways from the transient currents. This method has been used in studying lithium intercalation into graphite electrodes [5], where a momentary current increase is believed as a sign of nucleation and growth in the composing particles of the porous electrode. Allen et al. [6] first related the voltage-step responses of nano-LiFePO$_4$ electrodes with the Kolmogorov-Johnson-Mehl-Avrami (KJMA) equation. By fitting the "effective" rate constant, they found that the activation energy of surface reaction is lower than that of lithium diffusion in Li$_x$FePO$_4$ lattice. Okubo et al. [7] implemented similar analysis in studying nano-LiMn$_2$O$_4$ electrodes. More recently, Oyama et al. [8] adopted the first-order time-derivative of the KJMA equation to interpret the current responses of nano-LiFePO$_4$ porous electrodes at different voltage steps. By fitting the Avrami exponent, they concluded that at small voltage steps, nano-LiFePO$_4$ electrode undergoes one-dimensional (1D) two-phase reaction, while at large voltage steps, a non-equilibrium solid-solution pathway is possible, as predicted by theoretical studies of single nanoparticles [9, 10]. While all of these studies concluded 1D nucleation and growth mechanisms, none of them justified the fundamental basis of the KJMA equation with the physics of porous electrodes composed of countless individual nanoparticles.

According to the KJMA theory, randomly distributed germ nuclei in a large piece of metal/alloy will be activated to growth nuclei as a random process under a small temperature step, thus the transformed volume fraction can be modeled by the statistics of growth nuclei. One of the most important issues in this problem is the impingement of new phase growths. As shown in Fig. 1a, KJMA equation is based on the assumption that the actual increment of one growing grain of the new phase ($dV$) is on average the increment of the "extended volume" ($dV_{ext}$, assuming no impingement between growing grains) times the total fraction of the untransformed part, i.e. $dV=dV_{ext}(1-f)$ [11-14], where $f$ is the transformed fraction. However, new phase growths in porous electrodes are confined within individual composing particles. Even though the impingement of new phase growths would take place in single particles, it is physically unreasonable to link the microscopic increment to the global transformed fraction of the porous electrode, which has been proved by *electrochemical* experiments [15, 16] to be the *number fraction* of the transformed particles. If assume single nucleation event in each particle [15], the random nucleation process in a composite porous electrode is equivalent to the particle activation process. As shown in Fig. 1b, the porous electrode is composed of $FePO_4$ particles, phase-transforming (active) particles and $LiFePO_4$ particles. Each $FePO_4$ particle has a certain activation rate to become an active particle, into which lithium ions can insert continuously. And each active particle has another rate to be fully filled to $LiFePO_4$. Thus the discharging dynamics can be modeled as a special 3-state Markov chain [17].

By studying the population dynamics of phase-transforming nanoparticles, we developed a mathematical theory that can self-consistently explain the voltage-step responses of porous electrodes without modeling the microscopic phase transformation dynamics inside the composing nanoparticles. One of our most surprising findings is that the statistical kinetics of phase-transforming nanoparticles gives curves that could be misinterpreted as the nucleation and growth in single particles by the KJMA theory. Single particle behaviors concluded from experiments of porous electrodes are likely the results of statistical effects [3, 4].

## 2. Mathematical Models

Statistical effects in porous electrodes were seldom studied with mathematical models. For LiFePO$_4$ electrodes, Srinivasan and Newman [18] demonstrated that only with two different representative particles, could their model fit the experimental curves satisfactorily. Dreyer et al. [19] studied the discrete phase transformation in a many-particle system with an analogy to the interconnected balloon system, and conducted thermodynamic analysis to explain the voltage hysteresis of porous electrodes. For practical porous electrodes composed of countless nanoparticles, whether the particles are phase-transforming one by one or simultaneously can only be investigated by stochastic methods.

We begin by assuming a total of $N$ (normalized to 1) nanoparticles in a LiFePO$_4$ porous electrode are composed of $N_a(t)$ active particles, $N_t(t)$ transformed particles and $N_r(t)=1-N_a(t)-N_t(t)$ untransformed particles. We further assume that the activation rate for a continuous lithium insertion/extraction into each particle is $n(\Delta\phi,T,L)$, which may have the classic Arrhenius form [20] with $\Delta\phi$ the applied voltage, $T$ the temperature and $L$ the shape factor. The total increment of the population of the active particles can be written as,

$$dN_a(t) = \sum_{k=1}^{N_r(t)} n_k dt - dN_t(t) = \bar{n} N_r(t) dt - dN_t(t) \quad (1)$$

where $\bar{n}$ is the arithmetic average of the activation rates of all untransformed nanoparticles, and $dN_t$ is the incremental quantity of fully-transformed particles during time $dt$. Compared with Avrami's dynamic equations [12], $\bar{n}$ is equivalent to the probability of formation of active particles per untransformed particles per unit time.

2.1. Activation Rate

According to our dynamical study [9], there are two ways to activate continuous lithium insertion/extraction in single particles: (i) nucleation or spinodal decomposition triggered moving phase boundary, and (ii) homogenous reaction as a non-equilibrium quasi-solid solution. Previous experiment [15] suggested single nucleation event in each nanoparticle, thus the activation rate in case (i) is equal to the nucleation rate. Since nucleation in electrochemically driven system is accompanied by surface reaction, the applied voltage $\Delta\phi$ is critical for defining the nucleation barrier. Considering the thickness of the FePO$_4$/LiFePO$_4$ phase boundary [21, 22], we use the phase-field method [23, 24] to model the nucleation barrier as two parts,

$$\Delta G_n = \int_V \left[ \Delta g_n(c) + \kappa(\nabla c)^2 \right] dV \qquad (2)$$

where $c$ is the *local* mole fraction in a Li$_x$FePO$_4$ particle, $\Delta g_n(c) \equiv g_{sp} - g_h(c)$ is the *volume* energy barrier (Fig. 2a), and $\kappa(\nabla c)^2$ represents all sorts of energies induced by phase boundaries. For homogeneous states in metastable region, $\Delta g_n(c) > 0$, while for that in spinodal (unstable) region, $\Delta g_n(c) < 0$. In both cases, formation of phase boundary increases the nucleation barrier via $\kappa(\nabla c)^2$.

For an equilibrium particle without any applied voltage ($\Delta\phi=0$), the *volume* energy barrier to nucleation is $\Delta g_{sp}$, see Fig. 2a. Once a relatively small voltage is applied, e.g. $\Delta\phi<0$ for insertion, energy of the oxidized state of the surface reaction is elevated, resulting in a net insertion reaction (dashed curve in Fig.2b). Therefore, the concentration of solid-state lithium ions starts to grow till a new equilibrium state (point B in Fig 2c, and dash-dot line in Fig.2b). At this controlled equilibrium state, the *volume* energy barrier to nucleation is reduced to $\Delta g_n(c_B) = g_{sp} - g_h(c_B)$. Thus, $\Delta\phi$ serves as a thermodynamic constraint of the controlled equilibrium state by driving the dynamic process of lithium intercalation. Once a nucleation event occurs, the reaction equilibrium in the phase boundary region is broken, more lithium ions will be pumped in continuously, see Fig. 2d. Since the overpotential for surface reaction is much smaller than the nucleation barrier, it is very unlikely to have a second nucleation in single nanoparticles [15], making the random nucleation process equivalent to a particle activation process.

In situations of $\Delta G_n>0$, nucleation barrier will affect the nucleation rate via a general Arrhenius equation [20], e.g. $n = K\exp[-\Delta G_n(c,\Delta\phi)/RT]$, where $R$ is the gas constant, and the prefactor $K$ is the characteristic rate constant determined by material properties. In situations of $\Delta G_n=0$, nucleation no longer exists. $\bar{n}$ loses its microscopic basis of nucleation, and becomes the general activation rate, as case (ii). Therefore, if $\Delta\phi$ exceeds a critical value of order ~40mV [9, 10, 25], the activation rate becomes an average of the general activation rate of the particles undergoing homogeneous reaction and the averaged nucleation rate of the other ones, due to the potential distribution in the porous electrode. In both cases, the filling time of a $FePO_4$ nanoparticle is determined by surface reaction alone (or by diffusion for diffusion-limited cases [14]).

## 2.2. Particle-Filling Speed

Depending on the particle size, morphology, surface chemistry, electrochemical surroundings, etc., surface reaction rate could vary by orders, thus an estimation of particle-filling time larger than the diffusion time constant would be adequate for reaction-limited dynamics. For a model particle (similar to our nanomaterial, Fig. 4) with dimensions of 100nm×50nm×50nm ($a\times b\times c$), the diffusion time through $b$-channel at $D\approx10^{-8}cm^2/s$ is only 0.6ms [26]. We assume the filling time by reaction is $10^{-5}$ slower, which gives $\tau_{chn}\approx60s$ ($m_{chn}=0.017/s$) to fill a single channel. The number of $b$-channels of the model particle is calculated to be 20,000 with the spacing between $b$-channels ~0.5nm [27]. If the nanoparticle is filled one channel by one channel, the filling time of the particle is $\tau_{p,chn}\approx60s\times20,000=1,200,000s$ ($m_{p,chn}\approx10^{-7}/s$). If, otherwise, the particle is filled by a moving phase boundary, covering 200 channels, then $\tau_{p,bdr}=\tau_{p,chn}/200$ and $m_{p,brd}\approx10^{-4}/s$. Since the thickness of the phase boundary could span over 10nm due to the coherency strain [25], all channels of the nanoparticle will be active if the total length of the phase boundaries exceeds 500nm. Thus we have $m_{p,all}\approx10^{-3}/s$. Note that $m$ is simply an estimation of the filling speed. Different growth mechanism inside single particles could result in the same $m$. However, for a specific material, the above estimations define different regimes of the single particle dynamics: $m>10^{-3}/s$ represents all-channel-participating reaction in a single particle, and $10^{-7}/s<m<10^{-3}/s$ indicates typical moving phase boundary mechanism. These dynamic regimes can be easily calibrated for other materials. Similarly, the filling speed of a single nanoparticle can be defined as $m(\Delta\phi,T,L)$, so that the total increment of $N_t$ is,

$$dN_t(t) = \sum_{k=1}^{N_a(t)} m_k dt = \bar{m} N_a(t) dt \qquad (3)$$

where $\bar{m}$ is the averaged filling speed over all active nanoparticles, which can be understood as the probability to find a fully transformed particle per active nanoparticles per unit time. Parameters $\bar{n}$ and $\bar{m}$ allow our model to be applied to different materials with different kinds of growth mechanisms in single particles.

Without loss of generality, we can assume $N_a(0)=N_0$ to be the number of nanoparticles that have active phase boundaries at the beginning, and $N_t(0)=N_1$ the initial number of the transformed particles. Ordinary differential equations (1) and (3) can then be solved for $\bar{n} \neq \bar{m}$,

$$N_a = C_1 \exp(-\bar{m}t) + C_2 \exp(-\bar{n}t) \qquad (4)$$

$$N_t = 1 - C_1 \exp(-\bar{m}t) - \frac{m}{n} C_2 \exp(-\bar{n}t) \qquad (5)$$

where,

$$C_1 = \frac{N_0 \bar{m} + (N_1 - 1)\bar{n}}{\bar{m} - \bar{n}}, C_2 = \frac{(1 - N_1 - N_0)\bar{n}}{\bar{m} - \bar{n}} \qquad (6)$$

As shown in Fig. 3a, Eq. (4) gives similar curves produced by the first-order time-derivative of the KJMA equation [8]. For $\bar{n} > \bar{m}$, the population of active particles quickly reaches a peak and then decays exponentially. And for $\bar{n} < \bar{m}$, the increase of active particles is much slower and the population is relatively stable over a long period of time. While the non-monotonic curves generated by the KJMA equation are interpreted as the nucleation and growth mechanism in single composing particles, our equation demonstrates that those curves can simply be the results of the statistical kinetics of active particles.

2.3. Transient Current

Since active particles are the only ones carrying reaction current, the total current of the porous electrode is given by,

$$I = \sum_{k=1}^{N_a} m_k Q_k = \sum_{k=1}^{N_a} i_k = \bar{i} N_a \qquad (7)$$

where $Q$ is the lithium ion/vacancy capacity of an active nanoparticle, and $\bar{i}$ is the averaged reaction current over all active nanoparticles. For particles with uniform size, Eq.(7) can be approximated by $I \approx \bar{m} \bar{Q} N_a$, where $\bar{Q}$ is the arithmetic average of the distribution of $Q$ over all active particles. By taking $dI/dt=0$, we further find the time for the maximum transient current,

$$t_{\max} = \ln\left(-\frac{\bar{m}(N_0 \bar{m} + (N_1 - 1)\bar{n})}{\bar{n}^2 (1 - N_1 - N_0)}\right) \bigg/ (\bar{m} - \bar{n}) \qquad (8)$$

Since $t_{\max}$ depends on $N_0$ and $N_1$, the equilibrium status and the initial filling fraction of the porous electrode will affect the emergence and magnitude of the peak current. While $N_1$ can always be zero as long as we limit ourselves in a single voltage step, $N_0$ could be quite different, since porous electrodes may never reach the complete equilibrium [28].

In this section, we establish a simple model with two dynamic parameters, i.e. $\bar{n}$ and $\bar{m}$, which decouple the roles of random activation process and surface reaction during electrochemically driven phase transformation. $\bar{n}$ and $\bar{m}$ also provide a statistical explanation for the non-monotonic responses of porous electrodes, which have been misinterpreted by the shape factor, i.e. Avrami exponent, of the KJMA equation. To further study the transient currents of electrochemical systems, we still need $N_0$ and $\bar{Q}$ to provide the missing thermodynamic properties, which can be calculated theoretically. As shown in Fig. 3b, our model can accurately fit the experimental data of ultrathin porous electrodes developed by Oyama et al. [8].

3. **Experiments**

Nano-LiFePO$_4$ was synthesized in a Teflon-lined autoclave by a solvothermal method [29, 30]. The stoichiometric amounts of FeSO$_4$, LiOH and H$_3$PO$_4$ were dissolved in ethylene glycol and thoroughly mixed at room temperature. Then the dark green suspension was filled into the autoclave up to 90% volume. Maintained at 180 °C for 10 hours with a naturally cooling down to room temperature, some gray precipitates was obtained. After being washed with de-ionized water and ethanol, the precipitate was dried in vacuum at 80 °C for 12 hours. To further improve the electronic conductivity, the material was mixed with sucrose as the carbon-coating source (LiFePO$_4$:C=1:0.05, w/w), and then annealed in a tube furnace with flowing Ar. It was first kept at 200 °C for 1 hour and then heated to 650 °C at 5 °C/min and kept for 10 hours.

The final carbon-coated LiFePO$_4$ (LiFePO$_4$/C) and pristine LiFePO$_4$ were examined by X-Ray Powder Diffraction (XRD). Both XRD patterns in Fig. 4a can be indexed as the orthorhombic LiFePO$_4$ crystal (JCPDS Card No: 81-1173). And the Scanning Electron Microscope (SEM) image reveals that the LiFePO$_4$/C particles have a rather uniform size-distribution around 100nm, similar to our model particle.

LiFePO$_4$/C, PVdF and conductive carbon were weighted in ratio of 8:1:1, and mixed with some N-methylpyrrolidone (NMP) to make a smooth slurry, which was then coated onto an Aluminum foil and dried at 80 °C for 1 hour. Disk electrodes with diameter of 8mm were then punched off the cathode foil, and further dried in vacuum at 120 °C for 24 hours. CR2032 cells were assembled in Argon-filled glove box with Lithium metal as the anode, and 1M LiPF$_6$ in ethylene carbonate (EC)/dimethyl carbonate (DMC) (1:1 by volume) as the electrolyte. The cells were then cycled between 2.5V and 4.2V under different C-rates, shown in Fig. 4c. In each cathode, the mass of nano-LiFePO$_4$ is ~5mg. By assuming the same size of our model particle in section 2.2, and using the theoretical density 3.6g/cm$^3$ for LiFePO$_4$, the total number of nanoparticles is estimated to be $N\approx5.56\times10^{12}$, and the theoretical capacity of each particle is $Q=5.51\times10^{-7}$μAs, which gives the consistent capacity of the cathode as $NQ$=0.85mAh. Cells after two cycles of 0.1C charge/discharge were used in Potentiostatic Intermittent Titration Technique (PITT) experiments to find the phase-transformation voltage steps and the corresponding currents. The cutoff current was C/50 (17μA). Transient currents under different voltage steps are collected in Fig. 5.

## 4. Results

### 4.1. Small Voltage Steps

When relatively small voltage steps were applied (<40mV, Fig. 5a, b and c), all discharging curves exhibit linear decaying, but all charging curves develop plateaus. While this phenomenon may be considered as the general asymmetry between lithium insertion and extraction reported in many experiments [31, 32], our model suggests that it is a coupled result of asymmetries of nucleation barriers, surface reactions and initial status of the porous electrode, which can be captured by $\bar{n}$, $\bar{m}$ and $N_0$, respectively. Dynamically, the real nucleation barrier is a time-average of the controlled equilibrium barriers calculated by Eq. (2) during the transition from the initial equilibrium to the new equilibrium. Because this transition time is determined by the reaction rate, asymmetry between surface reactions will result in asymmetry between the dynamic nucleation barriers. As shown in Fig. 6a, curves with larger $\bar{n}$ and higher $N_0$ decay monotonically with quite linear beginnings. Because we assume $\bar{n}$ is constant throughout the process, $N_0$ serves as an approximation of the temporal variation of $\bar{n}$ at the beginning. Responses of $N_a$ exhibit the same trends as that of $I$ via Eq. (7).

Compared with the results from ultrathin electrodes [8], the tilted plateau and the turning point for a faster decaying in the responses of our thick electrode (Fig. 5) are very likely the result of transport effects [33, 34], which may complicate $\bar{n}$ and $\bar{m}$ with time and space dependence. However, our simple model can still capture the magnitudes of the peak currents and the decaying time approximately; therefore can be used to explain the behavior of thick electrodes qualitatively.

### 4.2. Large Voltage Steps

When relatively large voltage steps were applied (≥40mV, Fig. 5d, e and f), both charging and discharging currents develop tilted plateaus at the initial stage. However, a much more important feature is that the charging and discharging curves are getting closer (symmetric). At 150mV, the two transient currents decay to the cut-off current in one hour, and almost overlap with each other. These phenomena not only indicate the suppression of the asymmetry between charging and discharging, but also require higher $\bar{n}$ and $\bar{m}$. After $\bar{n}$ is increased to account for the high voltage, the only possibility for the process to finish sooner is increasing $\bar{m}$ (dash-dot curve in Fig. 6d), indicating an increase in numbers of active channels at the same time. As such, $\bar{m}$ becomes the typical filling speed of all-channel-participating reaction in single particles, which is consistent with the definition of quasi-solid solution [9] and the conclusion of Oyama et al. [8]. Although the dash-dot transient current decays monotonically from a very high value (Fig. 6d), the corresponding population of active particles does not change that much from the other control curves (Fig. 6e). The relatively low population of active particles indicates a discrete (particle-wise) phase transformation in the whole porous electrode, no matter the active particles are focused within a narrow region to form a discrete phase boundary or dispersed throughout the entire porous electrode. Only at extremely high $\bar{n}$, equivalent to very high overpotential, could all particles be activated simultaneously, as shown in Fig. 3a. In another word, under normal conditions with moderate overpotential, not all particles in the porous electrode are phase-transforming at the same time, even if the phase separation is suppressed in single particles. Therefore, dynamics of porous electrodes cannot be directly equated to that of a single composing particle. As long as the activation rate of untransformed nanoparticles (with $N_0 \ll 1$) is comparable to the mean filling speed of active nanoparticles (solid curves in Fig. 3a and Fig. 6d), non-monotonic transient currents will be produced. Since this feature comes from the random process, rather than the nucleation and growth mechanism in single particles, inferring phase transformation dynamics by the KJMA equation is invalid.

# 5. Discussion

## 5.1. Active Sites

As a generalization of our model, $N_a$ and $N_t$ can be treated as populations of active and transformed *sites*, rather than that of particles. Multiple-nucleation on active sites has been intensively studied in the field of electrodeposition. Fletcher [35] modeled crystal deposition and re-dissolution on active sites as a birth-and-death process. Milchev [36] proposed three rate constants to study the populations of the active sites and the nuclei. Bosco [37] studied the electrochemical phase formation process by formulating overlaps in terms of geometrical probability and nearest-neighbor statistics. D'Ajello et al. [38] coupled a diffusion equation with Fokker-Planck equation to model the transient current. However, active sites discussed in electrodeposition [14] are *static* tiny areas on the electrode surface that *can be activated* for nucleation, while that in our model are *dynamic* sites that *carrying reaction current*, i.e. the sites composing the reaction fronts (moving phase boundaries). Because of this subtle difference, our active sites within the fronts/boundaries have their own probability to be turned off as the impingement of grain growths, thus dismiss the difficulties introduced by "extended volume" and "overlap". This generalization allows our model to be applied to the other intercalation materials with different microscopic dynamics [39].

## 5.2. Phase Transformation Delay

Recently, *in situ* Powder Diffraction techniques are more and more used to investigate phase transformation of porous electrodes, which can distinguish different phases by detecting X-ray or neutron reflections from perfect crystalline lattices. However, in a working electrode, it is improbable to have fully relaxed lattice structures in active nanoparticles due to the diffusion of ions/polarons and the movement of phase boundaries. Thus the non-equilibrium new phases formed in the active nanoparticles are hard to be detected. The persistence of these active particles in porous electrodes will delay the emergence of the signal of the new phases, which until now has been attributed to surface amorphization [3] or solid solution reaction [4] in single composing particles. *In* *situ* powder diffraction experiments [3, 4, 31, 32, 40-43] inevitably mixed the information of single particle behavior with the statistical effects, where our model can help to decouple the physics.

Our model can be further generalized by including size-dependent material properties and porous electrode effects, such as temporal and spatial variations of overpotential and concentration. It provides complementary perspectives for battery research in terms of connecting the single particle physics [9, 25, 44] with the experimental results of porous electrodes. For electrodes composed of phase-separating particles, the population dynamics of active particles provides important implications for the homogenization of porous electrodes [33, 34, 45-47].

## 6. Conclusion

Based on the analysis of the nucleation barrier by phase-field method, the roles of nucleation and surface reaction in electrochemically driven phase-transformation processes have been decoupled for the first time, which allows for decoupling the activation rate and the filling speed from the "effective" rate constant of the KJMA equation and further modeling the statistical kinetics of the porous electrode as a simple Markov process among countless composing particles. We have demonstrated that the population dynamics of active nanoparticles exhibit non-monotonic transient currents, which could be misinterpreted as the nucleation and growth mechanism by the KJMA theory. Since not all particles are phase-transforming simultaneously under normal working conditions, electrode behavior should not be directly interpreted as single-particle dynamics. Instead of the Avrami exponent, the averaged filling speed $\bar{m}$ can be extracted from the voltage-step responses of porous electrodes to identify the phase transformation dynamics in single composing nanoparticles by comparing with the material-specific dynamic regimes.

## Appendix. Special Solutions of $N_a(t)$ and $N_t(t)$

For $\bar{n} = \bar{m}$, Eqs. (1) and (3) can be solved as,

$$N_a = C_1 \exp(-\bar{n}t) + C_2 t \exp(-\bar{n}t) \tag{A1}$$

$$N_t = 1 - \left(C_1 + \frac{C_2}{\bar{n}}\right)\exp(-\bar{n}t) - C_2 t \exp(-\bar{n}t) \tag{A2}$$

$$C_1 = N_0, C_2 = (1 - N_1 - N_0)\bar{n} \tag{A3}$$

For $N_a(t)=0$, Eqs. (1) and (3) will be reduced to

$$dN_t(t) = \bar{n}(1 - N_t(t))dt \tag{A4}$$

which can be easily solved to,

$$N_t(t) = 1 - \exp(-\bar{n}t) \tag{A5}$$

This equation is identical to the KJMA equation with Avrami exponent equals 1.

**Acknowledgement.** This work was supported by the Ministry of Science and Technology of China under a "973" Project (2011CB711202) and by a Tsinghua University Initiative Scientific Research Program (2011Z08139). The authors thank Dr. Xiankun Huang, Prof. Xiangming He and Prof. Jianbo Zhang for their help in experiments, and Prof. Martin Z. Bazant for his critical comments on an earlier version of the manuscript.

Figure 1

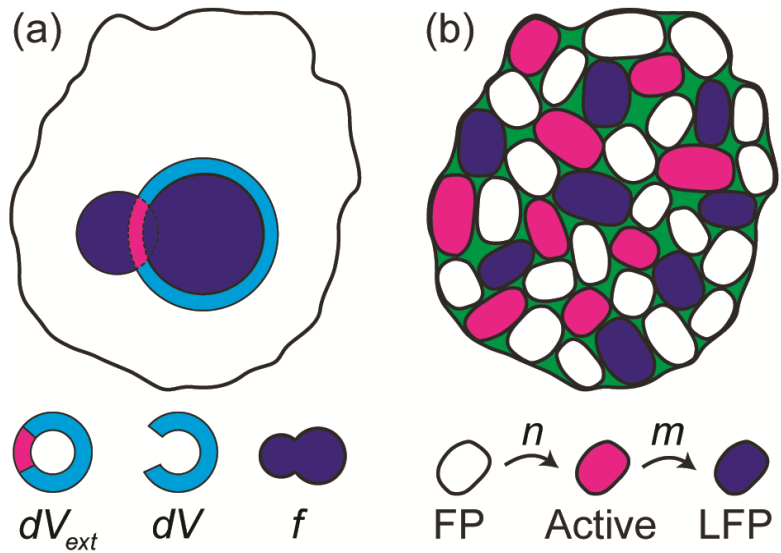

**Figure 1.** Schematic demonstrations of (a) nucleation and growth mechanism in a large piece of metal/alloy assumed by the KJMA theory and (b) stochastic process in a porous electrode composed of loosely contacted nanoparticles. FP stands for $FePO_4$ particle, which will transform to active particle at an activation rate of $n$. LFP stands for $LiFePO_4$ particle, which is transformed from the active particle at a filling speed of $m$.

Figure 2

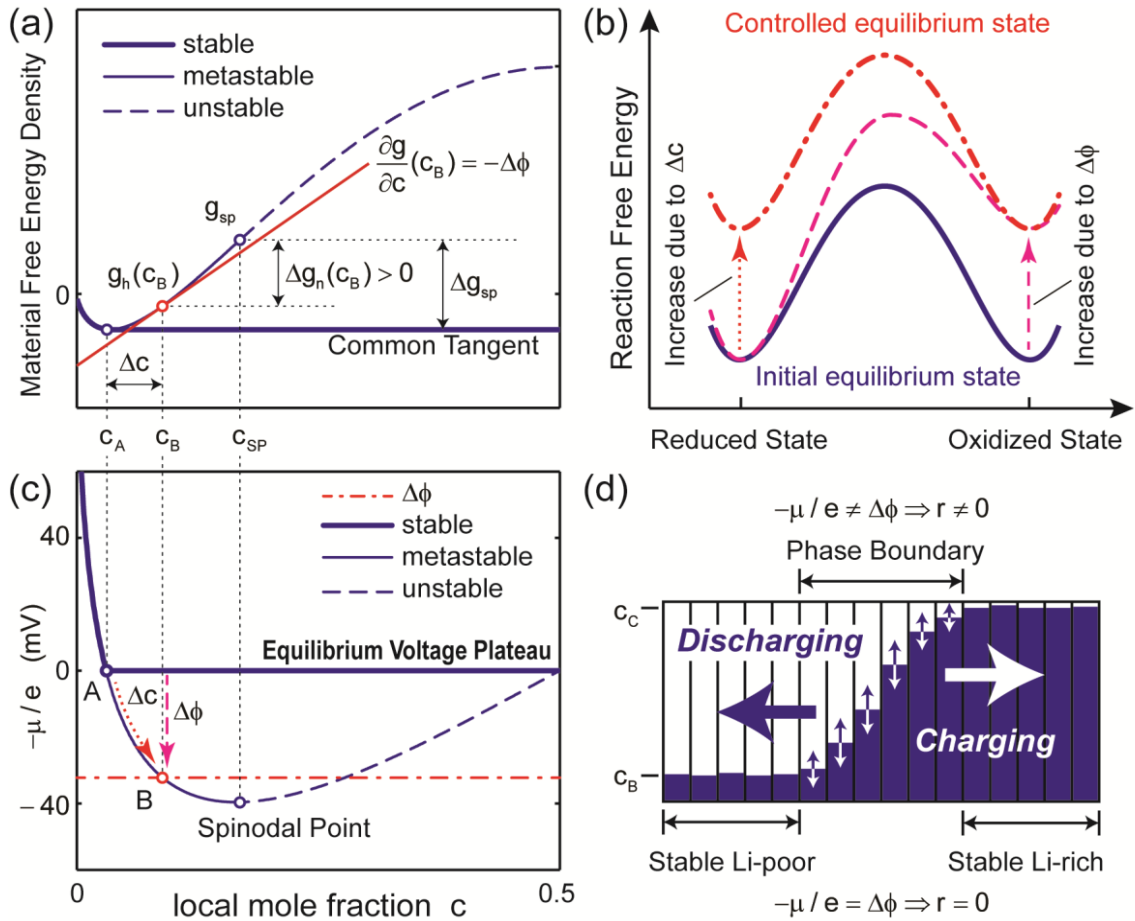

**Figure 2.** (a) Volume energy barrier at the controlled equilibrium state $c_B$; (b) dynamic process toward the voltage-controlled equilibrium state and (c) the corresponding response of the diffusional chemical potential $\mu=\partial g/\partial c$; (d) schematic demonstration of a Li$_x$FePO$_4$ particle with a moving phase boundary. $c_C$ is the stable Li-rich composition determined by the same $\Delta\phi$. Only in the phase boundary region, reaction rate $r\neq0$, could lithium ions be inserted/extracted, forming a moving phase boundary.

Figure 3

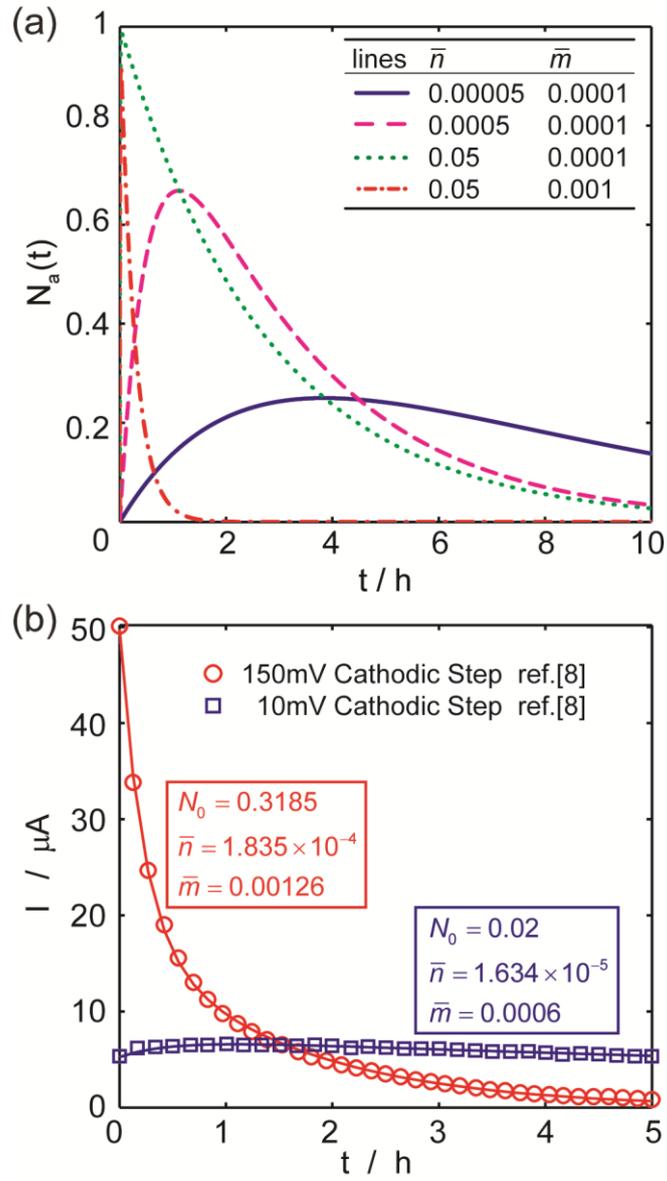

**Figure 3.** (a) Plot of $N_a(t)$ with $N_0=N_1=0$, and different $\bar{n}$ and $\bar{m}$. (b) Fitting results of the experimental data of ultrathin LiFePO$_4$ electrodes developed by Oyama et al.[8] The coefficients of determination are $R^2=0.965$ for 10mV step and $R^2=0.9995$ for 150mV step. Here, the electrode capacity was fitted, which in our experiments was calculated theoretically.

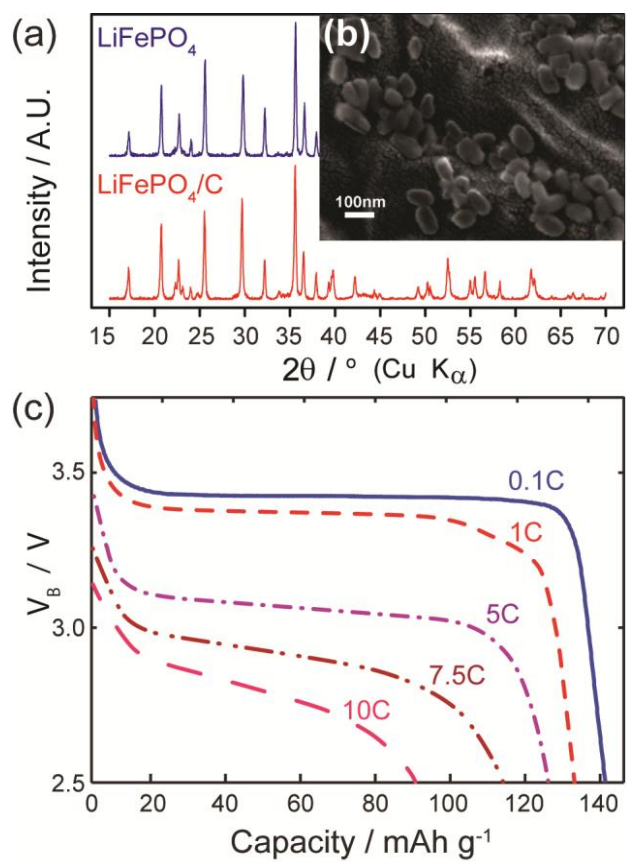

**Figure 4.** (a) XRD pattern (b) SEM image, and (c) discharging performance at different C-rates of LiFePO$_4$/C nanoparticles.

Figure 5

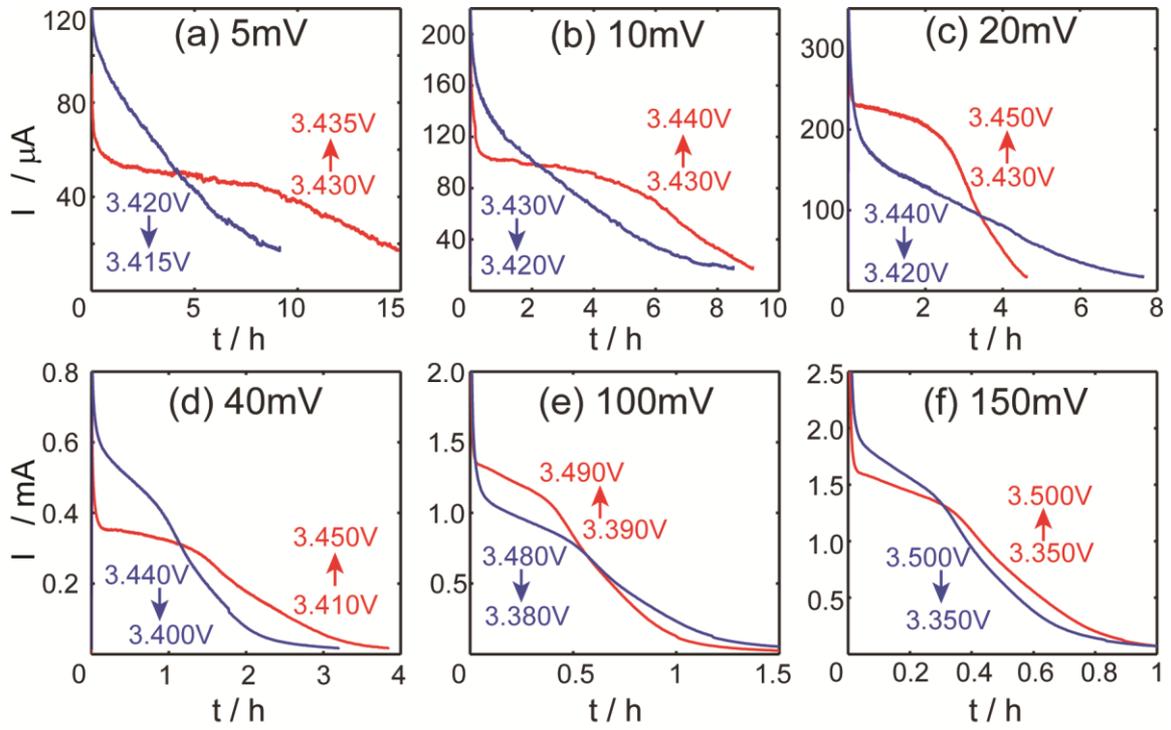

**Figure 5.** Transient currents under different voltage steps.

Figure 6

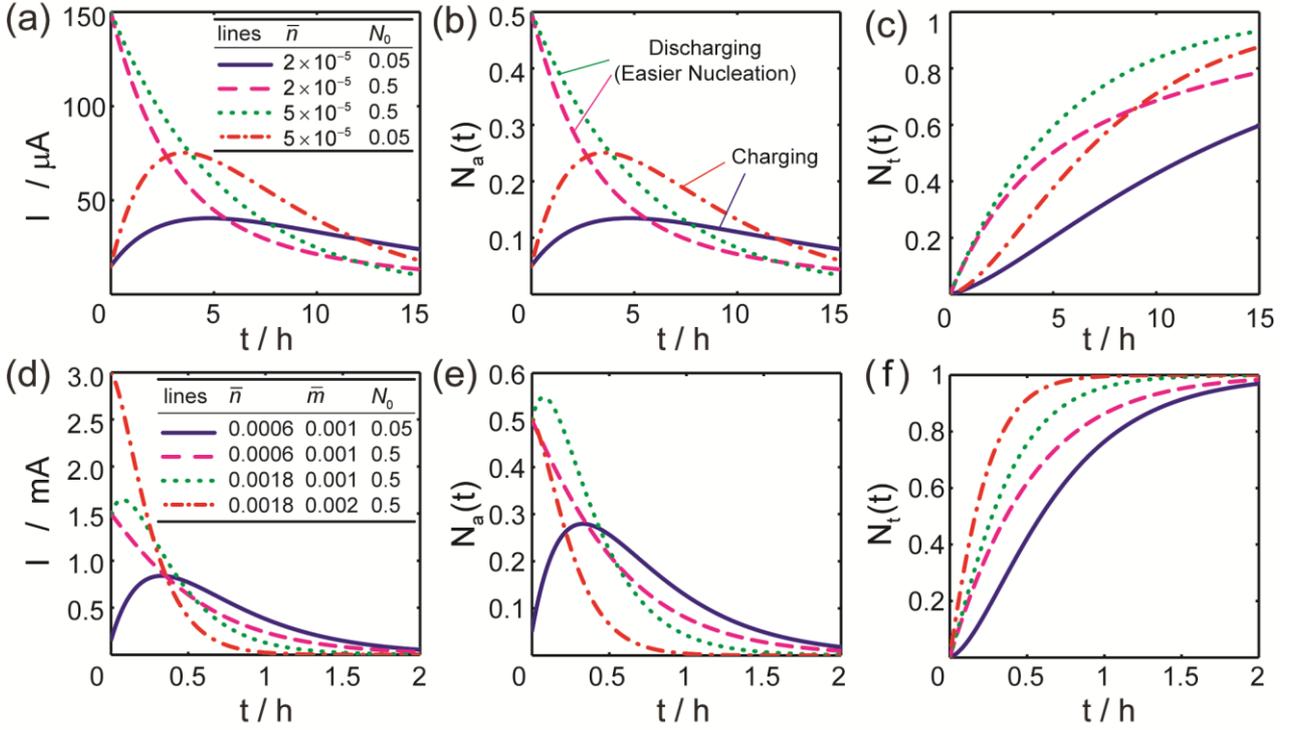

**Figure 6.** Theoretical responses of (a) the transient currents, populations of (b) active nanoparticles and (c) transformed particles under small voltage steps with $\bar{m}=10^{-4}$/s for typical moving phase boundary mechanism and $\bar{n} \approx 10^{-5}$/s guided by the results from Oyama et al. [8]; and (d) the transient currents, populations of (e) active nanoparticles and (f) transformed particles under large voltage steps with $\bar{n} \approx 10^{-3}$/s to account for high voltages and $\bar{m} \approx 10^{-3}$/s for fast reaction. For both cases, the theoretical capacity is 0.85mAh.